\documentclass[prl,twocolumn,superscriptaddress,letterpaper]{revtex4}
\usepackage{graphicx}
\begin{document}
\newtheorem{theorem}{Theorem}
\newtheorem{acknowledgement}[theorem]{Acknowledgement}
\newtheorem{algorithm}[theorem]{Algorithm}
\newtheorem{axiom}[theorem]{Axiom}
\newtheorem{claim}[theorem]{Claim}
\newtheorem{conclusion}[theorem]{Conclusion}
\newtheorem{condition}[theorem]{Condition}
\newtheorem{conjecture}[theorem]{Conjecture}
\newtheorem{corollary}[theorem]{Corollary}
\newtheorem{criterion}[theorem]{Criterion}
\newtheorem{definition}[theorem]{Definition}
\newtheorem{example}[theorem]{Example}
\newtheorem{exercise}[theorem]{Exercise}
\newtheorem{lemma}[theorem]{Lemma}
\newtheorem{notation}[theorem]{Notation}
\newtheorem{problem}[theorem]{Problem}
\newtheorem{proposition}[theorem]{Proposition}
\newtheorem{remark}[theorem]{Remark}
\newtheorem{solution}[theorem]{Solution}
\newtheorem{summary}[theorem]{Summary}   
\def\r{{\bf{r}}}
\def\i{{\bf{i}}}
\def\j{{\bf{j}}}
\def\m{{\bf{m}}}
\def\k{{\bf{k}}}
\def\kt{{\tilde{\k}}}
\def\mt{{\hat{t}}}
\def\mG{{\hat{G}}}
\def\mg{{\hat{g}}}
\def\mGa{{\hat{\Gamma}}}
\def\mS{{\hat{\Sigma}}}
\def\mT{{\hat{T}}}
\def\K{{\bf{K}}}
\def\P{{\bf{P}}}
\def\q{{\bf{q}}}
\def\Q{{\bf{Q}}}
\def\p{{\bf{p}}}
\def\x{{\bf{x}}}
\def\X{{\bf{X}}}
\def\Y{{\bf{Y}}}
\def\F{{\bf{F}}}
\def\G{{\bf{G}}}
\def\bG{{\bar{G}}}
\def\mbG{{\hat{\bar{G}}}}
\def\M{{\bf{M}}}
\def\V{\cal V}
\def\tchi{\tilde{\chi}}
\def\tx{\tilde{\bf{x}}}
\def\tk{\tilde{\bf{k}}}
\def\tK{\tilde{\bf{K}}}
\def\tq{\tilde{\bf{q}}}
\def\tQ{\tilde{\bf{Q}}}
\def\si{\sigma}
\def\ep{\epsilon}
\def\hep{{\hat{\epsilon}}}
\def\al{\alpha}
\def\be{\beta}
\def\ep{\epsilon}
\def\bep{\bar{\epsilon}_\K}
\def\up{\uparrow}
\def\de{\delta}
\def\De{\Delta}
\def\up{\uparrow}
\def\dwn{\downarrow}
\def\ksi{\xi}
\def\etha{\eta}
\def\product{\prod}
\def\goto{\rightarrow}
\def\switch{\leftrightarrow}

\title{Ferromagnetic Spin Coupling as the Origin of $0.7$
Anomaly in Quantum Point Contacts} 

\author{K.~Aryanpour and J. E. Han}
\affiliation{Department of Physics, SUNY at Buffalo, 
Buffalo, NY 14260} 

\date{\today}
\begin{abstract}

We study one-dimensional itinerant electron models with ferromagnetic
coupling to investigate the origin of $0.7$ anomaly in quantum point
contacts. Linear conductance calculations from the quantum Monte Carlo
technique for spin interactions of different spatial range suggest
that $0.7(2e^{2}/h)$ anomaly results from a strong interaction of
low-density conduction electrons to ferromagnetic fluctuations formed
across the potential barrier. The conductance plateau appears due to the
strong incoherent scattering at high temperature when the
electron traversal time matches the time scale of dynamic ferromagnetic
excitations.

\end{abstract}
\pacs{}
\maketitle 

Quantum point contacts (QPC) are narrow constrictions inside
two-dimensional electron gas. They construct one of the building blocks
of submicrometer devices such as quantum dots and
qubits~\cite{lantz,thomas}. The dc conductance through a QPC is quantized in
steps of $G_{0}=2e^{2}/h$ \cite{wees,wharam}. However, experiments also
reveal the appearance of an additional shoulder in the conductance
measurement near $0.7G_{0}$ widely referred to as the $0.7$
anomaly \cite{thomas}. The origin of $0.7$ anomaly in QPC has remained
a puzzle over almost a decade.  The evolution of the $0.7G_{0}$ plateau
to $0.5G_{0}$ with magnetic field and the enhancement of the
$g$ factor~\cite{thomas} have strongly suggested that the origin of the
anomaly is the electron spin. 

A number of scenarios have been proposed, such as spin polarization of
the itinerant electrons
\cite{kui,calmels,zabala,bruus,reilly,starikov,tokura}, ferromagnetic
correlation~\cite{spivak,bartosch,yang}, formation of a spin $1/2$
magnetic moment in the conductance
channel~\cite{rejec,cronenwett,hirose,puller} and Kondo
effect~\cite{meir,cronenwett,hirose},
Hubbard chain~\cite{kirchner,alvarez,louis,schmeltzer,lunde,sloggett,syljuasen},
Wigner crystallization and antiferromagnetism~\cite{matveev}.  These
approaches have produced a wide range of different phenomenologies,
sometimes inconsistent with experiments, and there is no widely accepted
microscopic theory to date. The problem is partly due to the approximate
methods used in the strongly interacting limit and therefore it becomes
essential to perform exact calculations to test microscopic models
against experiments. Here we use numerically accurate
quantum Monte Carlo technique to study the strong
correlation effects in QPC devices.

We find that the 0.7 anomaly at high temperature arises from the
incoherent electron scattering from itinerant ferromagnetic fluctuations
near the Stoner instability~\cite{doniach}, in the strong correlation
limit of low electron density created by spatially inhomogeneous gate
potential. We show, through a comparison with a model
with on-site interactions, that the relevant electron
scattering is due to the spin fluctuations which are spatially coherent
across the potential barrier. With decreasing temperature, the magnetic
excitation becomes slower than the itinerant electrons. The current is
then carried by the quasiparticles and the 0.7 plateau gradually
disappears. With the Zeeman magnetic field, the 0.7 plateau evolves to a robust 0.5
plateau in agreement with experiments.

\begin{figure}
\includegraphics[width=3.1in]{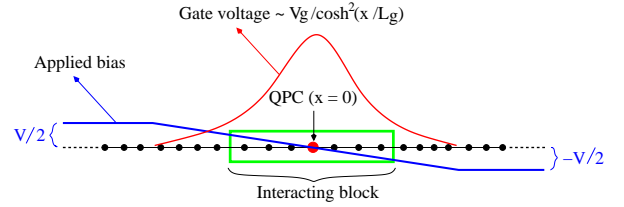}
\caption[a] {(color online) Profile of the 1D chain with the interacting
block (including seven sites in this figure). The
gate voltage potential $V(x)$ acts as an adiabatic potential
barrier through the QPC region. A bias with a ramp passing
through the QPC is applied across the chain to compute the dc
conductance in the linear response regime.}
\label{model}
\end{figure}

We model our system using a one-dimensional (1D) electron gas with spin-spin interaction
among itinerant electrons as depicted in Fig.\ref{model} with the
assumption that the higher 1D subbands play indirect roles for the
first conductance plateau. The Hamiltonian reads
\begin{eqnarray} \label{eq:1D-spin-contm} {\cal H} & = & \int dx \
\psi^{\dag}(x)\left[-\frac{\hbar^2}{2m}\frac{\partial^2}{{\partial
x}^2}+V(x)-\mu+\frac{1}{2}\vec{\sigma}\cdot\vec{H}\right]\psi(x)
\nonumber \\ & & + \int dx \left\{K_{1}(x)\left[\vec{s}(x)\right]^2 + \
\frac12 K_{2}(x)\left[\partial_{x}\vec{s}(x)\right]^2 \right\}\,,
\end{eqnarray} with $\psi(x)=[\psi_{\uparrow}(x),\psi_{\downarrow}(x)]^{T}$ the field operator vector, $\mu$ the chemical
potential, $H$ the Zeeman magnetic field, $\vec\sigma$ the Pauli matrices
and $V(x)$ the external gate voltage barrier in order to pinch off
the electron current through the QPC. $V(x)$ is defined in our model as
$V(x)=V_{g}/\cosh^2(x/L_{g})$ with $x=0$ corresponding to the center of
the chain, $L_{g}$ a characteristic length and $V_{g}$ the gate voltage.
The operator $\vec{s}(x)=\psi^{\dag}(x)\frac{\vec{\sigma}}{2}\psi(x)$
represents the spin density of itinerant electrons along the chain.
Spins interact locally with the coupling constant
$K_{1}(x)=\alpha(x)K_{1}$ ($K_1<0$ for repulsive on-site Coulomb
interaction) with $\alpha(x)=1/\cosh(x/L_{s})$ an attenuating function
with characteristic length $L_{s}$. We set the spin coupling to
adiabatically fall off to take into account the screening effects in the
leads and to reduce the backscattering due to interaction away from the
QPC constriction.  $K_{2}(x)=\alpha(x)K_{2}$ $(K_2>0)$ is the
coefficient of the gradient term accounting for the ferromagnetic
Heisenberg interaction. We discretize the continuum model Hamiltonian to
a tight-binding chain of lattice constant $\Delta x$ with the
nearest-neighbor hopping $t$.  Defining $t=\frac{\hbar^2}{2m{\Delta
x}^2}$, $\mu=\mu-\frac{\hbar^2}{m{\Delta x}^2}$, $\sqrt{\Delta x}\
\psi(x_{i})=c_{i}$, $J_{0}= -(\frac{K_{1}}{\Delta
x}+\frac{K_{2}}{{\Delta x}^3})$, $J_{1}=\frac{K_{2}}{{\Delta x}^3}$ and
$\vec{s_{p}}=c^{\dag}_{p}\frac{\vec{\sigma}}{2}c_{p}$ the discretized
Hamiltonian reads  
\begin{eqnarray} \label{eq:1D-spin-disc} {\cal
H}=-t\sum_{<ij>,\sigma}c^{\dag}_{i\sigma}c_{j\sigma}
-\sum_{i\sigma}\left(\mu- V_{i}+\frac{1}{2}\sigma H\right)
c^{\dag}_{i\sigma}c_{i\sigma} \nonumber \\ -\sum_{p\in
block}(J_{0}\alpha^2_{p}\vec{s}_{p}\cdot\vec{s}_{p}
+J_{1}\alpha_{p}\alpha_{p+1}\vec{s}_{p}\cdot\vec{s}_{p+1}) \,,
\end{eqnarray}
with $H$ taken along the $z$ direction and $\sigma=\pm 1$ the spin
index. The microscopic parameters $K_1$ and $K_2$ are unknown and we
treat $J_{0}, J_1>0$ (for ferromagnetic coupling) as the model
parameters throughout this letter. The index $p$ runs only within the
interacting block in Fig.\ref{model} near the QPC saddle point. The
discretization is valid since we are in the low-density limit with
$\langle c^\dagger_{i\sigma}c_{i\sigma}\rangle <1$ inside the
interacting block.  Near the pinch-off gate voltage, $\langle
c^\dagger_{i\sigma}c_{i\sigma}\rangle$ at the top of the potential
barrier rapidly approached zero in the following calculations. Using
$m\approx0.067m_{e}$ for GaAs and $\Delta x\approx 20~\textrm{nm}$
(experimental QPC length is about $200\ \textrm{nm}$, roughly $10\Delta
x$ for $L_{g}=4$), $t=1.4~\textrm{meV}$.  Due to CPU limitations, we restrict the interacting block to about seven sites.

\begin{figure}
\includegraphics[width=3.4in]{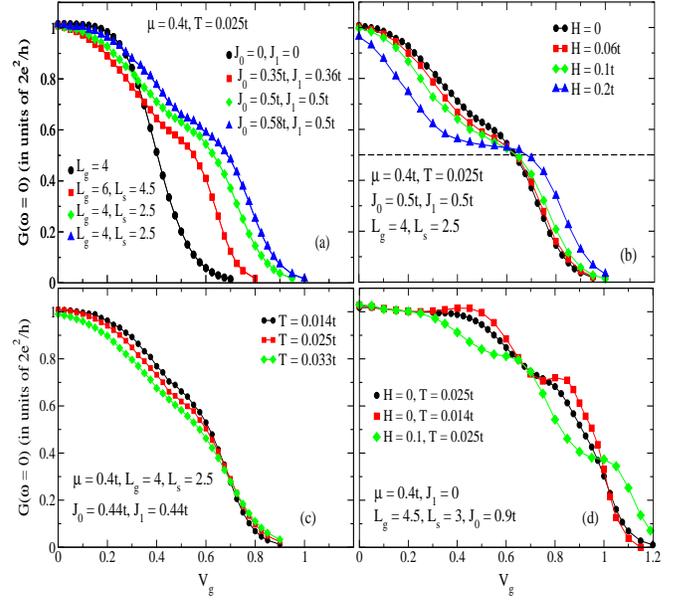}
\caption[a]{(color online) (a) dc conductance as a function of 
gate voltage $V_{g}$ at different values for $L_{g}$, $L_{s}$, $J_0$ and
$J_1$ at $T = 0.025t$. Conductance plateaus form at $0.6-0.7$ times of
$G_0=2e^2/h$. (b) The
evolution of the plateau to $0.5G_{0}$ with the Zeeman
magnetic field. (c) The gradual increase of conductance with
decreasing temperature. (d) Purely local model $(J_1=0)$ as a function
of temperature and the Zeeman magnetic field. Local interaction produces
qualitatively different transport mechanism from the nonlocal model.}
\label{anomaly}
\end{figure}

In our calculations, we modify
Eq.(\ref{eq:1D-spin-disc}) by allowing the nonlocal part of the
interaction term to extend beyond nearest neighboring sites. We define
the block spin operator $\vec{S}=\sum_{p\in block}\alpha_{p}\vec{s}_{p}$
and rewrite a new Hamiltonian
\begin{eqnarray}
\label{eq:1D-spin-modif}
{\cal H}=-t\sum_{<ij>,\sigma}c^{\dag}_{i\sigma}c_{j\sigma} 
-\sum_{i\sigma}\left(\mu- V_{i}+\frac{1}{2}\sigma
H\right)c^{\dag}_{i\sigma}c_{i\sigma} \nonumber \\ 
-\sum_{p\in block}\left(J_{0}-\frac{J_{1}}{2}\right)\alpha^2_{p}\vec{s}_{p}
\cdot\vec{s}_{p}-\frac{J_{1}}{2}\vec{S}\cdot\vec{S} \,.
\end{eqnarray}
Compared to Eq.(\ref{eq:1D-spin-disc}),
Eq.(\ref{eq:1D-spin-modif}) incorporates stronger spin interaction among
all the spins within the interacting block.
This modification makes the decoupling scheme in quantum Monte Carlo more efficient.
Interactions beyond nearest neighbors can be thought of a
coarse-grained effective Hamiltonian on the
discretized lattice in the low wave-vector limit. The effective
interaction
results from virtual fluctuations to high momentum states which are
excluded in the discretized model and it takes a form similar to the
RKKY interaction. Since we are interested in the low-density limit near
the pinch-off regime with the effective Fermi wave-vector $k_{\rm
F,eff}$
inside the constriction approaching zero, the
$k_{\rm F,eff}R_i$ factor for position $R_i$ inside the constriction
also goes to zero and the
effective spin interaction over the interacting block becomes
predominantly ferromagnetic.

We use a continuous Hubbard-Stratonovich decoupling
for the $\vec{S}\cdot\vec{S}$ term in Eq.(\ref{eq:1D-spin-modif})
and discrete decoupling for local $\vec{s}_{m}\cdot\vec{s}_{m}$
term. We calculate the dc conductance using the Kubo formula in the
linear response regime, $G_{\mathrm dc}(\omega=0)=\lim_{\omega \rightarrow 0}
{\textrm Re}~i \int_{\scriptscriptstyle
0}^{\scriptscriptstyle\infty}e^{i \omega
t'}\big<\big[j(t'),H_{\mathrm sd}\big]\big>dt'\,,$
where $j(t')=i et
\sum_{\sigma}[c_{1,\sigma}^{\dagger}(t')c_{0,\sigma}(t')-c_{0,\sigma}^{\dagger}(t')c_{1,\sigma}(t')]$
is the current operator evaluated at the center of the QPC and
$H_{\mathrm sd}=-e\sum_{m,\sigma}V(x_m)n_{m,\sigma}$ is the external perturbation
across the chain with $V(x_m)$ the normalized source-drain bias with
maximum (minimum) voltage $1/2$ ($-1/2$) on the left (right) hand side
as depicted in Fig.\ref{model}. The range of summation in $H_{\mathrm sd}$,
$|m|\alt 100$, produced well-converged conductance.
Conductance by the Kubo formula is
obtained in terms of the bosonic Matsubara frequencies
($i\nu_{n}=i\frac{2\pi n}{\beta}$ where $\beta=1/k_{B}T$ with $k_{B}$ the
Boltzmann constant and positive integer $n$) and it needs to be
analytically continued
to real frequency to take the dc limit $G_{\mathrm dc}(\omega=0)$. This task is
done by fitting the conductance defined on the Matsubara frequency into the Lehman representation
$G(i\nu_{n})=i\int_{\scriptscriptstyle-\infty}^{\scriptscriptstyle\infty}d\omega'\frac{\rho(\omega')}{i\nu_{n}-\omega'}$
with the spectral function $\rho(\omega)$ as the fitting parameter. This
method has been extensively tested to an excellent agreement in
comparison with the rational function
fit~\cite{syljuasen}.
After taking the analytic continuation
$i\nu_{n}\rightarrow\omega+i\eta$, we obtain the conductance
$G_{\mathrm dc}(\omega=0)=\rho(0)$.

\begin{figure}
\includegraphics[width=2.5in]{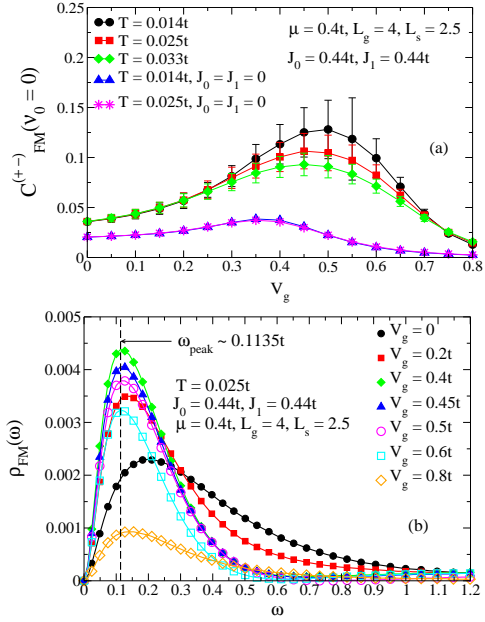}
\caption[a]{(color online) (a) Static ferromagnetic spin 
correlation function versus the gate voltage
at different temperatures. The susceptibility has a strong
many-body enhancement (compared to the noninteracting model). (b)
Spectral function for dynamic ferromagnetic spin susceptibility
$\rho_{\rm FM}(\omega)$
at different values of gate voltage at
$T = 0.025t$. The dashed line indicates the excitation
frequency.}
\label{chi.0.7anomaly}
\end{figure}

Fig.\ref{anomaly}(a) plots the dc conductance as a function of the gate
voltage $V_{g}$ for different values of $L_{g}$ and $L_{s}$ at several
$J_{0}$ and $J_{1}$ values and fixed chemical potential $\mu=0.4t$ when
$H=0$. With increasing interaction strength, the plateau evolves from
near $0.5G_0$~\cite{sloggett} to higher values $0.6-0.7G_0$ in the
strongly interacting limit with $J_0,J_1>\mu$.  Fig.\ref{anomaly}(b)
exhibits the gradual evolution of the plateau near $0.7G_{0}$ to
$0.5G_{0}$ as the Zeeman magnetic field is applied. The significantly
wide gate voltage interval for the plateau region ($\Delta V_{g}\approx
0.4t$ at $0.5G_{0}$ for $H=0.2t$) compared to the bare Zeeman splitting
($\Delta V_{g}=H=0.2t$) is a clear indication for the enhanced
$g$ factor as seen in the experiment~\cite{thomas}. Fig.\ref{anomaly}(c)
shows that, with decreasing temperature, the conductance plateau
consistently moves upward with decreasing width.
Despite the phase problem of the quantum Monte Carlo method for temperature $T<0.014t$, we are
able to capture the correct trend as seen in the experiment for
$0.014t\le T \le 0.033t$.  $T=0.025t$ corresponds to
$T\approx0.41\textrm{K}$, falling within the experimental
range~\cite{cronenwett}. Features in Fig.\ref{anomaly}(a)-(c) correctly
reproduce the experimental results on the temperature and magnetic field
dependence.

Fig.\ref{anomaly}(d) plots the conductance in the purely local
limit $(J_{1}=0)$. The local spin model is equivalent to the repulsive
Hubbard model~\cite{kirchner,alvarez,louis,syljuasen} with the on-site
Coulomb parameter $U$ given as $U=3J_0/4$
by redefining the gate potential $V_i$ to absorb the one-body terms.
Although the local limit produces a
well-defined 0.7 feature at zero field, it is inconsistent with the 0.7
phenomenology.
First, the 0.7 feature becomes more pronounced as temperature is
lowered. Second, more interestingly, at finite magnetic field two
plateaus 
appear with the 0.7 feature shifted to higher conductance and another
plateau emerging near $G\sim 0.3G_0$. It is very interesting that these
results are consistent with the scenario of spin singlet-triplet
formation 
discussed in
Refs.\cite{rejec2,puller}. The main difference here is that the spin is
self-generated from itinerant electrons in our model, not as an external
spin~\cite{puller} or from a quasibound state~\cite{rejec2}. Near the
pinch-off, electron and spin densities are low and the spin-singlet does not
form and the conductance plateau does not appear at $H=0$.
However, at finite $H$, the spin moment becomes enhanced enough
to produce the spin-singlet conductance plateau.

In the presence of nonlocal interactions, an itinerant electron
interacts with many
neighboring sites and the resulting spin multiplets are not necessarily
$S=0$ or $S=1$. With the finite interacting range, a spatially coherent
ferromagnetic state extends over the interacting
block at finite $H$ and becomes harder to be flipped by electron
scattering. Therefore, the nonlocal interaction blocks
the minority-spin band and the 0.5 plateau results from spin splitting,
instead of the 
$0.3$ plateau through the singlet formation. It has been shown
previously that ferromagnetic coupling beyond local interaction
stabilizes the ferromagnetic phase in a uniform 1D chain~\cite{yang}.

\begin{figure}
\includegraphics[width=3.4in]{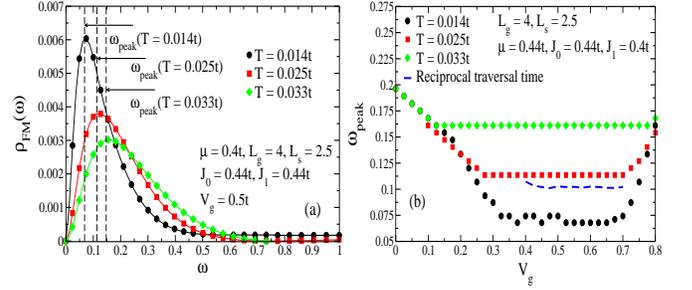}
\caption[a]{
(color online) (a) $\rho_{\rm FM}(\omega)$ at $V_{g} = 0.5t$ for
different
temperatures. The peak frequency $\omega_{peak}$ corresponds to
ferromagnetic excitation energy. (b) The excitation 
energy as a function of gate voltage $V_{g}$ at different
temperatures. When the traversal time $\tau_{tr}$ (dashed line) is
long inside the constriction, \textit{i.e.} $\omega_{peak}\tau_{tr}>1$, the
electron scattering becomes strong and the conductance plateau results.
}
\label{magnon-energy}
\end{figure} 

Due to the low electron density within the
constriction at the pinch-off, the ferromagnetic
spin correlations are considerably enhanced at low $T$. We
compute the ferromagnetic spin correlation function (FSCF),
$C^{(+-)}_{FM}(\nu_{n})=\frac{1}{N^2}\sum_{p,m\in
block}\int_{\scriptscriptstyle 0}^{\scriptscriptstyle\beta}e^{i
\nu_{n}\tau}\big<S^{+}_{p}(\tau)S^{-}_{m}(0)\big>d\tau\,,$
with $\nu_{n}=\frac{2\pi n}{\beta}$ ($n \ge 0$) the bosonic Matsubara
frequency, $\tau$ the imaginary time and $N$ the number of sites inside
the interacting block in Fig.\ref{model}. Fig.\ref{chi.0.7anomaly}(a)
plots the static FSCF $(\nu_{0}=0)$ as a function of
$V_{g}$ at different $T$ values corresponding to the conductance curves
in Fig.\ref{anomaly}(c). In comparison with the noninteracting
system, the static FSCF is significantly enhanced as $T$ is lowered when
there is interaction. Enhancement of the static FSCF is
effective only around the pinch-off due to the
singular nature of the 1D density of states. The gate voltage
range for the maximum enhancement of static FSCF also coincides with
that of the plateaus in Fig.\ref{anomaly}(a)-(c), indicating the effect
of strong ferromagnetic correlations on the appearance of the plateau.
The dynamic ferromagnetic spin susceptibility can be obtained by
analytically continuing the FSCF using the same method  employed for the
conductance. Through the fluctuation-dissipation theorem,
$\rho_{FM}(\omega)$, the spectral function for magnetic excitations can
be obtained from the analytically continued
$C^{(+-)}_{FM}(i\nu_{n}\rightarrow \omega+i\eta)$ from positive $\nu_n$. In
Fig.\ref{chi.0.7anomaly}(b), $\rho_{FM}(\omega)$ has been plotted for
different values of $V_{g}$ at $T=0.025t$ corresponding to
Fig.\ref{chi.0.7anomaly}(a). The height of the peak grows as $V_{g}$
increases towards the pinch off, reaching its maximum for $V_{g}\approx
0.4t-0.5t$, around $\omega_{peak}\approx 0.1135t$ as indicated by the
dashed line. As $V_{g}$ is further increased, the height of the peak
decreases in agreement to the decrease in the static FSCF in
Fig.~\ref{chi.0.7anomaly}(a).
However, the location of the peak continues to stay at
$\omega_{peak}\approx 0.1135t$ as the characteristic excitation energy
at $T=0.025$ up to $V_{g}\approx0.7t$ where ferromagnetic correlations
have almost been obliterated as shown in Fig.\ref{chi.0.7anomaly}(a).

Tokura and Khaetskii~\cite{tokura} addressed the effect of
scattering for electrons in the current due to the local paramagnons as
the origin of the $0.7$ anomaly in the QPC using the second order
perturbation approach with local interactions.
Following B\"uttiker and Landauer in Ref.\cite{buttiker}, we argue that
magnons, perceived as spin waves with characteristic frequency
$\omega_{peak}$, can strongly interfere with electrons while
transmitting through the gate voltage barrier at frequencies
$\omega_{peak}\sim 1/\tau_{tr}$ where $\tau_{tr}$ is the traversal time for
tunneling through the barrier in the absence of magnons defined as
$\int_{\scriptscriptstyle x_{1}}^{\scriptscriptstyle
x_{2}}dx\sqrt{\frac{m}{2[V(x)-E]}}$, with $x_{1}$ and $x_{2}$ the
classical turning points, $E\approx\mu$ the incident energy and
$V(x)=V_{g}/\cosh^2(x/L_{g})$ the gate voltage barrier.

Fig.\ref{magnon-energy}(a) presents $\rho_{FM}(\omega)$ for $V_{g}=0.5t$
(maximum FSCF in Fig.\ref{chi.0.7anomaly}(a)) at three different $T$
values. The characteristic frequency $\omega_{peak}$ decreases as the
temperature is lowered and therefore the spin waves soften at the onset
of Stoner instability in the system as explained in detail in
Ref.\cite{doniach}.  Fig.\ref{magnon-energy}(b) plots $\omega_{peak}$ as
function of $V_{g}$ for these three different temperature values along
with the reciprocal traversal time $1/\tau_{tr}$ (nearly constant as a
function of $V_{g}$). While $\omega_{peak}\sim 1/\tau_{tr}$, strong
interference between the spin fluctuations and electrons tunneling
through the gate voltage barrier leads to the suppression of the
conductance for the gate voltage values in the vicinity of the pinch
off. The manifestation of this suppression is the appearance of a
plateau in the conductance as a function of the gate voltage near the
pinch off. With decreasing temperature and therefore $\omega_{peak}\ll
1/\tau_{tr}$, as seen in Fig.\ref{magnon-energy}(a) and (b) for $T=0.014t$,
the itinerant electrons begin to see an effectively static ferromagnetic
mean field and scattering of electrons off magnetic excitations becomes
coherent. As a result, the conductance increases and the plateau
disappears.

We have demonstrated the incoherent electron scattering due to
ferromagnetic spin-spin correlations in a quasi-1D chain as the
underlying physics behind $0.7$ anomaly phenomenon in the QPC.
We thank Jonathan Bird and Igor \v Zuti\'c
for helpful discussions. This project was supported by NSF DMR-0426826
and we acknowledge the CCR at the SUNY Buffalo for computational
resources.

\end{document}